\documentclass[aps,prl,showpacs,twocolumn]{revtex4}
\usepackage{graphicx}
\begin{document}

%%%%%%%%%%%%%%%%%%%%%%%%%%%%%%%%%%%%%%%%%%%%%%%%%%%%%%%%%%%%%%%%%%%%%%%%%%%%%%%%%%%
\newcommand{\figureheight}{8.2 cm}
\newcommand{\putfig}[2]{\begin{figure}[h]
        \special{isoscale #1.bmp, \the\hsize \figureheight}
        \vspace{\figureheight}
        \caption{#2}
        \label{fig:#1}
        \end{figure}}

        % almost universal commands for equations and references
\newcommand{\eqn}[1]{(\ref{#1})}

\newcommand{\be}{\begin{equation}}
\newcommand{\ee}{\end{equation}}
\newcommand{\bea}{\begin{eqnarray}}
\newcommand{\eea}{\end{eqnarray}}
\newcommand{\bean}{\begin{eqnarray*}}
\newcommand{\eean}{\end{eqnarray*}}

\newcommand{\nn}{\nonumber}
%%%%%%%%%%%%%%%%%%%%%%%%%%%%%%%%%%%%%%%%%%%%%%%%%%%%%%%%%%%%%%%%%%%%%%%%%%%%%%%%%%%

%%% ----------------------------------------------------------------------

%%versione del
%%%%%%%%%%%%%%%%%%

%%%%%%%%%%%%%%%%%%

\title{ Integer Spin Hall Effect in Ballistic Quantum Wires}
\author{S. Bellucci $^a$ and P. Onorato $^a$ $^b$ \\}
\address{
        $^a$INFN, Laboratori Nazionali di Frascati,
        P.O. Box 13, 00044 Frascati, Italy. \\
        $^b$Dipartimento di Scienze Fisiche,
        Universit\`{a} di Roma Tre, Via della Vasca Navale 84,
00146 Roma, Italy} %\maketitle \widetext
\date{\today}
%\maketitle
%\widetext
\begin{abstract}

We investigate the ballistic electron  transport in a two
dimensional Quantum Wire under the action of an electric field
($E_y$). We demonstrate how the presence of a Spin Orbit coupling,
due to the uniform electric confinement field gives a
non-commutative effect as in the presence of a transverse magnetic
field.

We discuss how the non commutation implies an edge localization of
the currents depending on the electron spins also giving a
semi-classical spin dependent Hall current.

We also discuss how it is possible obtain a quantized Spin Hall
conductance in the ballistic transport regime by developing the
Landauer formalism and show the  coupling between the  spin
magnetic momentum and the orbital one due to the presence of a
circulating current.

\end{abstract}


\maketitle

{\it INTRODUCTION-} In the last decade  spin-dependent transport
phenomena have attracted  a lot of interest because of their
potential for future electronic device applications. It follows
that the electrical control of spins  in nanostructures is of
basic interest and has great potential in semiconductor
electronics {\it "spintronic"}\cite{spintro,wolf}. Since
1990\cite{Datta} it was discussed  how the electrical field can be
used to modulate the current and the essential role, which the
field-dependent Spin Orbit (SO) coupling plays in this mechanism,
was shown. Recently many works \cite{[1–5]} have been devoted to
the study of injection of spin-polarized charge flows into the
nonmagnetic semiconductors from ferromagnetic metals.

\

Nevertheless the SO interaction has an essentially relativistic
nature it can also give rise to some sensible  effects on the
semiconductor band structure\cite{Stormer,Nitta}. In low
dimensional semiconductor devices, as Quantum Dots\cite{noidot}
and Quantum Wires (QWs), a natural SO coupling is always present
which arises due to structural inversion asymmetry in quantum
heterostructures\cite{Kelly} where two-dimensional (2D) electron
systems are realized. In this case the mechanism of the SO
interaction originating from the interface field is known as
Rashba effect, because it was first introduced by
Rashba~\cite{Rashba}.

\

The recent developments in the analysis of SO effects have open a
new field of research oriented toward the phenomenology of the so
called Spin Hall Effect (SHE).
 In 1999, Hirsch\cite{Hir} proposed that when a charge current
circulates in a paramagnetic metal a transverse spin imbalance
will be generated, giving rise to what he called  "spin Hall
voltage". Recent discovery of intrinsic spin-Hall effect in
p-doped semiconductors by Murakami et al. \cite{[6]} and in Rashba
spin-orbit (SO) coupled two-dimensional electron system (2DES) by
Sinova et al. \cite{[7]} may possibly lead to a new solution to
the issue. In last years  Raimondi and Schwab calculated the
spin-Hall conductivity for a two-dimensional electron gas  varying
the strength and type of disorder\cite{rs}. The theory of
transport in the presence of SO interaction including disorder was
developed
%\cite{[1415]}
 also in the presence of a magnetic field: the Rashba effect in the presence of an
in-plane magnetic field yields a characteristic anisotropic
conductivity as a function of the magnetic field\cite{[1415]}. The
effects of SO coupling were also investigated in the ballistic
regime for QWs\cite{governale,morozb,me} and, in a recent letter
the presence of a Mesoscopic SHE was predicted also in a
Multiprobe SO Coupled Semiconductor Bridges in the Ballistic
regime\cite{Mpb}.

\

The Hall effect occurs when an electric current flows through a
conductor in a magnetic field, creating a measurable transverse
voltage. On a fundamental level, this effect originates because
the magnetic field exerts a force on the moving charge carriers,
which pushes them to one side of the conductor. The resulting
buildup of charge at the sides of the conductor ultimately
balances this magnetic field- induced force, producing a
measurable voltage between opposite sides of the conductor. In
analogy to the conventional Hall effect, the SHE has been proposed
to occur in paramagnetic systems as a result of spin-orbit
interaction, and refers to the generation of a pure spin current
transverse to an applied electric field even in the absence of
applied magnetic fields. A pure spin current can be thought of as
a combination of a current of spin-up electrons in one direction
and a current of spin-down electrons in the opposite direction,
resulting in a flow of spin angular momentum with no net charge
current. Similar to the charge accumulation at the sample edges,
which causes a Hall voltage in the conventional Hall effect, spin
accumulation is expected at the sample edges in the SHE. In a
recent article Kato et al. detected an electrically induced
electron-spin polarization near the edges of a semiconductor
channel  and imaged with the use of Kerr rotation microscopy. The
polarization is out-of-plane and has opposite sign for the two
edges, consistent with the predictions of the spin Hall
effect\cite{kato}.

\

 Here we discuss the case of a quasi one dimensional clean
 QW, first by analyzing the conventional Integer Quantum Hall Effect (IQHE)
 in the presence of a transverse magnetic field ($B$),
 then by discussing the case of SHE, for $B=0$.
 In this theoretical approach we neglect the effect of the Rashba coupling and
 we just take in account the electric fields acting in the plane where the electron are confined to move.

Thus we start from an introduction of the model, then we discuss
how the quantized transverse conductance ($G_{x-y} \equiv G_H$),
corresponding to
 the  IQHE, can
 be easily calculated  in QWs  starting from the Landauer
 formula.
Then we demonstrate a formal analogy between the model of QW in
the presence of a transverse magnetic field  and the one where
dominates the SO coupling: thus we extend our results to the SHE.

\

{\it MODEL-} Semiconductor QWs are quasi 1D devices   of width
less than $1000 \AA$\cite{thor} and length of some microns  (here
we think to a QW where $L_x \sim  30-100 nm$, $L\sim 10-100 \mu
m$,$L_z\lesssim 10 nm$). In these devices, where the electron
waves are in some ways analogous to electromagnetic waves in
waveguides,  the electrons are confined to a narrow quasi one
dimensional channel with motion perpendicular to the channel
quantum mechanically frozen out. Such wires can be fabricated
using modern semiconductor technologies such as electron beam
lithography and cleaved edge overgrowth.

From a theoretical point of view a QW is usually defined by a
parabolic confining potential along one of the directions in the
plane\cite{me}: $V_W(x)=\frac{m_e}{2}\omega_d^2 x^2$.

\

{\it SINGLE SPINLESS PARTICLE IN A MAGNETIC FIELD -} Here we
summarize some known results following refs.\cite{me,me1} and
bibliography therein. We consider a uniform magnetic field $B$
along the $\hat{z}$ direction acting on the QW and we choose
the gauge  ${\bf A}=(0,Bx,0)$. %It follows  that the single  particle Hamiltonian is
%\begin{equation}\label{hnw}
%H = m_e\frac{\bf v^2}{2}+\frac{m_e\omega_d^2}{2}x^2
%\end{equation}
%where $m_e v_y=p_y-eBx/(m_e c)$ and $m_e v_x=p_x$.
Now  we introduce the cyclotron frequency $\omega_c=\frac{eB}{m_e
c}$, the total frequency $\omega_T=\sqrt{\omega_d^2+\omega_c^2}$
and  ${\bf \pi}\equiv\left\{{\bf p}-\frac{e}{c}{\bf A}({\bf
R})\right\} $. Because of $\left[H,p_y\right]=0$ we can write
\begin{equation}\label{hnw2}
H =\frac{\pi_x^2+\pi_y^2}{2m_e}+V_W(x)=
\frac{\omega_d^2}{\omega_T^2}\frac{p_y^2}{2m_e}+\frac{p_x^2}{2m_e
}+\frac{m \omega_T^2}{2}(x-x_0)^2,
\end{equation}
where $x_0=\frac{\omega_c p_y}{\omega_T^2 m_e}$
%.
%\
%
%From a classical point of view, the  solution of the Hamilton
%equations, derived from eq.(\ref{hnw2}), gives \bea\label{eq0x}
%x(t)&=&x_0+R\cos(\omega_T t +\varphi_0) \\
%y(t)&=&y_0- \frac{\omega_T}{\omega_c} R \sin(\omega_T t
%+\varphi_0)+ v_d t, \label{eq0y}\eea where
and the {\it drift velocity},  $v_d$, is $v_d=%\frac{\partial H}{\partial p_y}=
\frac{ \omega_d^2p_y}{\omega_T^2m_e}$. It follows
that, in the presence of magnetic field along $z$,  two electrons,
moving along the $y$ direction with opposite versus (i.e. $\pm
p_y\rightarrow \pm v_d$), are localized on the two opposite edges
($\pm x_0$) . Thus the states corresponding to these localized
currents are also known in quantum mechanics as {\it edge
states}\cite{bvh}. The edge localization could also be seen as a
consequence of  the commutation properties, \bea
\left[\pi_x,\pi_y\right]=i \hbar m_e \omega_c ,\eea and of the
presence of a confinement potential (i.e. $V_W(x)$).

\

From the Quantum Mechanical point of view, the diagonalization of
the hamiltonian in eq.(\ref{hnw2}) gives two terms, i.e. a
quantized harmonic oscillator ($n$ labels the subband) and a
quadratic free particle-like dispersion ($p_y=\hbar k$)
\bea\label{enk} \varepsilon_{n,k}=\frac{\omega_d^2}{2m_e
\omega_T^2}\hbar^2 k^2+\hbar \omega_T(n+\frac{1}{2}), \eea

This kind of factorization does not reflect itself in the
separation of the motion along each axis because the shift in the
center of oscillations along $x$ depends on the momentum
$p_y=\hbar k_y$. From eq.(\ref{enk}) it follows the Fermi
wavevector as
$$
k_F(\varepsilon_F,\omega_c,n)=\sqrt{\frac{2m_e \omega_T^2}{\hbar^2
\omega_d^2}\left(\varepsilon_F-\hbar
\omega_T(n+\frac{1}{2})\right)}.
$$
Next we say than the $n-th$ subband is open if, after fixing the
Fermi energy ($\varepsilon_F$), results $k_F$ real (i.e.
$\left(\varepsilon_F-\hbar \omega_T(n+\frac{1}{2})\right)>0$). The
number of open subbands is labeled by $N_s$.

 \

The presence of an uniform electric field along the $y$ direction
localized in  the stripe $a/2>y>-a/2$ can be introduced as a
potential \bea\label{Vy} V(x,y)= E_y\, y\;\; \vartheta\left(a^2-4
y^2\right)+ E_y\, a\;\; \vartheta\left(y-a/2\right),\eea where
$\vartheta\left(x\right)$ is the Heaveside step function and
$\Delta V=E_y a$ can be assumed as a small bias Voltage
difference.

In the stripe where the electric field does not vanish, the
classical solution of the Hamilton equations yields
\bea\label{eqmx}
x(t)&=&x_0+R\cos(\omega_T t +\varphi_0) + v_{H}t \\
y(t)&=&y_0- \frac{\omega_T}{\omega_c} R \sin(\omega_T t
+\varphi_0)+ v_d t - \frac{1}{2}a_y t^2 , \eea where  the {\it
Hall velocity}, $v_{H}$, is $v_{H}=%\frac{\partial H}{\partial p_x}=
\frac{\omega_c E_y e }{\omega_T^2m_e}$, while $a_y=\frac{E_y e
}{m_e}$.

\

{\it BALLISTIC CONDUCTANCE -} In the regime of ballistic
transport the scattering with impurities can be neglected, because
both the width  and the length of the QW  are much larger than the
mean free path $\ell$. In this regime the Landauer formula
allows one to write the conductance in terms of transmission probabilities
of propagating modes at the Fermi level\cite{bvh}.

Next we consider a QW attached to two reservoirs at $y=\pm \infty$
with a current injected at $y=-\infty$.
  Scattering
within the QW, mainly due to the presence of the electric field
from eq.(\ref{Vy}), may reflect part of the injected current back
into the bottom reservoir. If we limit ourselves to a fixed
subband, $n$, a fraction $T_n$  of the injected current  $J_n$ is
transmitted to the reservoir at the top. Then the corresponding
diffusion current in the QW reads $j_y^+(n) \propto |t(k_n)|^2 v_d e
/L$, where $|t(k_n)|^2=T_n$ is the transmission coefficient.

 The
density of the states is obtained from eq.(\ref{enk}):
$$d\nu(k_n)= g_s \frac{\omega_T^2}{\omega_d^2}\frac{L}{2 \pi} dk_n,$$
where we introduce $g_s=2$ corresponding to the spin degeneration.
It could be shown  that the states which contribute to the
transport, for the $n-th$ subband,  have an energy
$\varepsilon_F+e\Delta V
>\varepsilon>\varepsilon_F$, with $\varepsilon_F$ the Fermi
energy, we have \bea I_{y}^+(n)&=&\int |t(k_n)|^2 \frac{v_d e }{
L}d\nu(k_n) \nonumber \\ &=&
\int_{\varepsilon_F}^{\varepsilon_F+e\Delta V}|t(k_n)|^2\frac{g_s
e}{h} d\varepsilon  \nonumber \\ &\approx&
|t(k_n(\varepsilon_F))|^2\frac{g_s e^2}{h}\Delta V. \eea

It is trivial to calculate  the transmission coefficient obtained  by
considering  the scattering potential in eq.(\ref{Vy})
$$|t(k_n(\varepsilon_F))|=\vartheta\left(\varepsilon_F-\hbar \omega_T(n+\frac{1}{2})-E_y
a \right)$$ It follows the longitudinal conductance, according the
Landauer formula,  \bea \label{gyy}
G_{y-y}=\sum_{n=0}\frac{I_y^+(n)}{\Delta V}=\sum_{n=0}
|t(k_F(n)|\frac{2 e^2}{h}=\frac{2 e^2}{h} N_t ,\eea where $ N_t$
is the number of open subbands   in the asymptotic region
($y>a/2$).

\

 From the localization of the edge
states now we can  deduce that a transverse current have to appear
in the stripe where the electric field does not vanish. This
current, $I_H(n)$, is due to the presence of an Hall velocity,  as
shown in eq.(\ref{eqmx}), and it gives a contribution of a
quantum of conductance to $G_H$ just if $|t|^2=0$ and
$\varepsilon_F>\varepsilon_{n,0}$.

We now introduce $I_0=(2e^2 \Delta V)/h$ as the current quantum,
and then we apply  the continuity equation for the currents.
Following the schematic representation in Fig.(1),
 $I_y^0= N_s I_{0}$ is the
injected current, which is localized on the right side of the QW,
$I_y^+=N_t I_{0}$ is the current measured at the top end of the QW
and it is localized on the right edge too, it follows that a
reflected current is present ($I_y^0=I_y^++I_H$)
$I_H=(N_s-N_t)I_{0}$ and from the discussed localization we know
that it is localized on the left side of the QW in the asymptotic
region $y<-a/2$. Thus in the stripe $-a/2<y<a/2$ there should  be
a current, $I_H$ in the $x$ direction from right to left. So we
obtain \bea \label{gxy}G_{y-x}=G_{H}=\sum_{n=0}
|r(k_n(\varepsilon_F)|\frac{2 e^2}{h}=\frac{2 e^2}{h}(N_s- N_t)
.\eea Now we can conclude that IQHE can be explained in terms of
transmitted and reflected channels in a QW.

In Fig.(1) we show a schematic behaviour of the electrons in the
QW when a magnetic field is present also focusing on the presence
of rings of current corresponding to the  reflection from the
electric field barrier.

 %%%%%%%%%%%%%%%%%%%%%%%%%%%%%%%%%%%%%%%%%%%%%%%%%%%%%%%%%%%%%%%%%%%%%%%%%%%
\begin{figure}
\includegraphics*[width=1.0\linewidth]{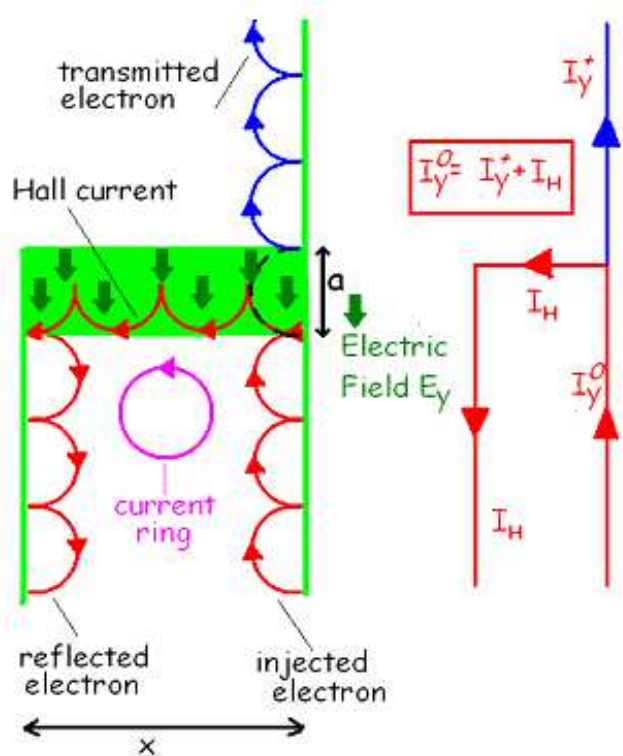}
\caption{(Color online){Schematic behaviour of currents in the QW.
}}
\end{figure}
%%%%%%%%%%%%%%%%%%%%%%%%%%%%%%%%%%%%%%%%%%%%%%%%%%%%%%%%%%%%%%%%%%%%%%%%%%%%
%%%%%%%%%%%%%%%%%%%%%%%%%%%%%%%%%%%%%%%%%%%%%%%%%%%%%%%%%%%%%%%%%
The conductance in eq.(\ref{gxy}) is shown in Fig.(2.left) where
we represent the $G$ as a function of the strength of the electric
field.

 %%%%%%%%%%%%%%%%%%%%%%%%%%%%%%%%%%%%%%%%%%%%%%%%%%%%%%%%%%%%%%%%%%%%%%%%%%%
\begin{figure}
\includegraphics*[width=1.0\linewidth]{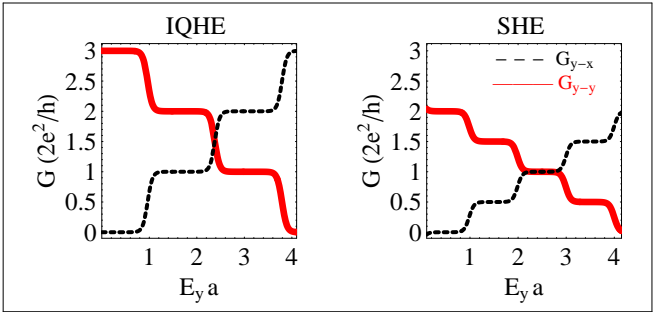}
\caption{(Color online){Longitudinal and Hall conductance for IQHE
and SHE: on
 the left integer hall effect,  for a fixed value of the magnetic
field ($\omega_c=\omega_d$); on the right SHE. }}
\end{figure}
%%%%%%%%%%%%%%%%%%%%%%%%%%%%%%%%%%%%%%%%%%%%%%%%%%%%%%%%%%%%%%%%%%%%%%%%%%%%
%%%%%%%%%%%%%%%%%%%%%%%%%%%%%%%%%%%%%%%%%%%%%%%%%%%%%%%%%%%%%%%%%

 \

{\it INTEGER QUANTUM SPIN HALL EFFECT -} Now we can extend our
calculations to the case where no magnetic field is present but
introducing the effect of the Spin Orbit coupling.

An electron moving in an electric field experiences not only an
electrostatic force but also a relativistic influence due to the
SO interaction.
 This manifests itself  in an interaction
term in the hamiltonian which couples the in-plane electron
momentum with the electron spin.

The SO interaction comes from the expansion quadratic in $v/c$ of
%eq.(\ref{H_d})
Dirac equation~\cite{Thankappan} and  is due to the Pauli coupling
between the spin momentum of an electron and a magnetic field,
which appears in the rest frame of the electron, due to its motion
in the electric field. { It follows} that the effects of an
electric field (${\bf E}({\bf R})$ where ${\bf R}$ is the 3D
position vector) on a moving electron have to be analyzed starting
from the following hamiltonian\cite{morozb}:
\begin{equation}
\hat H_{SO} = -\frac{\hbar}{(2m c)^2}\;{\bf E}({\bf R})
\left[\hat{{\bf \sigma}}\times \left\{\hat{\bf p}-\frac{e}{c}{\bf
A}({\bf R})\right\}\right]. \label{H_SO}
\end{equation}
Here $m$ is the free electron mass, %$\frac{\partial}{\partial x^\mu}$ are the canonical momentum operators,
 $\hat{{\sigma}}$ are
the Pauli matrices, ${\bf A}$ is the vector potential and we
introduce $\alpha\equiv \frac{\hbar^2}{(2m c)^2}$.

Next we take in account just electric field in the plane where the
QW lies. This hypothesis is quite different from the usual
treatment of the Rashba coupling in semiconducting devices, that
we discussed in a previous paper\cite{me} and will analyze in the
future.

The starting point is the Hamiltonian of one electron in the QW
where we introduce the SO term in eq.(\ref{H_SO}). In our case we
can consider the electric field due to eq.(\ref{Vy}) but also the
one corresponding to the harmonic confinement ($E_x(x)=-m_e
\omega_d^2 x$), thus \bea\label{hnws} H = \frac{p_x^2+p_y^2}{2m_e
}+\frac{m \omega_d^2}{2}x^2 + \frac{e E_y \hbar}{(2m_ec)^2}
\left[\sigma_x p_z -\sigma_z p_x \right] \nonumber \\ + \frac{m_e
\omega_d^2 x \hbar}{(2m_ec)^2} \left[\sigma_z p_y -\sigma_y p_z
\right]. \eea Now we can consider that the degree of freedom
corresponding to $z$ is frozen out because  the ratio between the
energies of the confined states along the different directions,
$x$ and $z$, is  $\varepsilon_z/\varepsilon_x>>10$, then we can
assume $\langle p_z\rangle=0$. Thus in what follows we neglect the
term with $p_z$ ($[H,\sigma_z]=0$) and introduce
$\pi_x=p_x-\sqrt{U_x} \sigma_z$ (with $\sqrt{U_x}=\frac{e E_y
\hbar m_e}{(2m_ec)^2}$), $\pi_y=p_y-m_e\Omega_c x \sigma_z$ (with
$\Omega_c=\frac{ \omega_d^2\hbar}{m_e(2 c)^2} $). These new
momenta correspond to the commutation properties: \bea
\left[\pi_x,\pi_y\right]=i \hbar m_e \Omega_c \sigma_z .\eea Thus
we can write
\begin{equation}\label{hnws2}
H =\frac{\pi_x^2+\pi_y^2}{2m_e}+V_W(x)-U_x -\frac{m_e
\Omega_c^2}{2}x^2 ,
\end{equation}
and then  we introduce the new constants
$\Omega_d^2=\omega_d^2-\Omega_c^2$ and the total frequency
$\Omega_T=\sqrt{\Omega_d^2+\Omega_c^2}$ so that eq.(\ref{hnws2})
becomes
\begin{equation}\label{hnws3}
H =
\frac{\Omega_d^2}{\Omega_T^2}\frac{p_y^2}{2m_e}+\frac{p_x^2}{2m_e
}+\frac{m \Omega_T^2}{2}(x-X_0)^2-U_x,
\end{equation}
where $X_0=s \frac{\Omega_c p_y}{\Omega_T^2 m_e}$ and $s=\pm1$
corresponds to the spin polarization along the $z$ direction.

From the discussed formal analogy, it emerges the presence of a
Spin Hall velocity $$v_{H}=%\frac{\partial H}{\partial p_x}=
s\frac{\Omega_c E_y e }{\Omega_T^2m_e},$$ clearly depending on
the spin polarization.

 \
 %%%%%%%%%%%%%%%%%%%%%%%%%%%%%%%%%%%%%%%%%%%%%%%%%%%%%%%%%%%%%%%%%%%%%%%%%%%
\begin{figure}
\includegraphics*[width=1.0\linewidth]{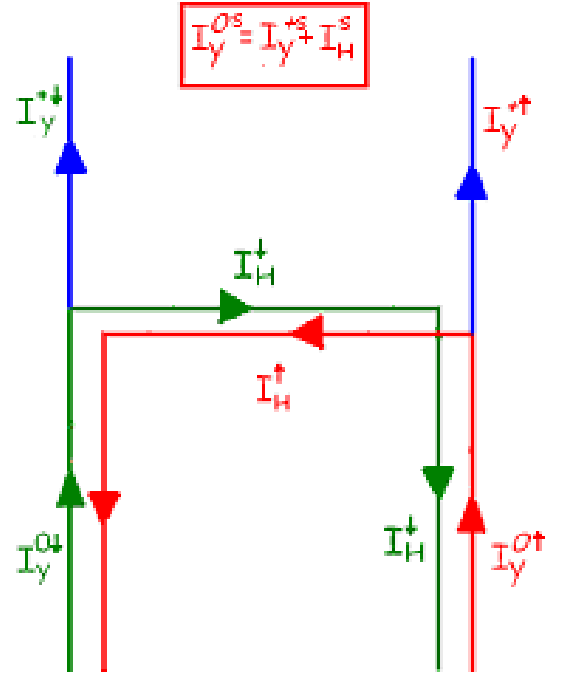}
\caption{(Color online){Schematic representation of the spin
currents in the QW. }}
\end{figure}
%%%%%%%%%%%%%%%%%%%%%%%%%%%%%%%%%%%%%%%%%%%%%%%%%%%%%%%%%%%%%%%%%%%%%%%%%%%%
%%%%%%%%%%%%%%%%%%%%%%%%%%%%%%%%%%%%%%%%%%%%%%%%%%%%%%%%%%%%%%%%%
 Following the schematic representation in Fig.(3.top),
 $I_y^{0,\uparrow}= N_s I_{0}/2$ is the
injected current, which is localized on the right-hand side of the
QW, $I_y^{+,\uparrow}=N_t I_{0}$ is the current measured at the
top end of the QW, and it is localized on the right-hand edge too.
Hence, it follows that a reflected current is present
($I_y^{0,\uparrow}=I_y^{+,\uparrow}+I_H^{\uparrow}$)
$I_H^{\uparrow}=(N_s-N_t)I_{0}/2$, and from the discussed
localization we now that it is localized on the left-hand side of
the QW in the asymptotic region $y<-a/2$. Thus, in the stripe
$-a/2<y<a/2$ there should  be a current, $I_H^{\uparrow}$ in the
$x$ direction, from right to left. If we assume that a spin
polarized current is injected in our device (e.g. because we
consider ferromagnetic leads) the presence of the plateaux in the
longitudinal conductance depending on the strength of the electric
field ($G_{y-y}=N_t e^2/h$ due to $g_s=1$) can be also read as the
presence of a transverse Spin Hall current with a quantized
conductance ($G_{x-y}=(N_s-N_t) e^2/h$). This is represented in
Fig.(2.right).

\

When we take into account a spin unpolarized current it is clear
that $I_y^{+}=I_y^{+,\uparrow}+I_y^{+,\downarrow}$, which gives the
conductance in the form of eq.(\ref{gyy}). The symmetry of the
device implies that the charge Hall current vanishes,
$I_H=I_H^{\uparrow}+I_H^{\downarrow}=0$. In this case we can
define also the spin Hall current as $$
I_{sH}=I_{H}^{\uparrow}-I_H^{\downarrow},$$ whence it follows that
  \bea\label{gsh}
G^e_{sH}=\frac{I_H^\uparrow-I_H^\downarrow}{\Delta V}=
\frac{2e^2}{h}(N_s-N_t). \eea It could be very interesting to
observe that a spin current, linked to a vanishing charge current,
is present everywhere on the edge of the wire, so that we can
define some {\it spin edge states} analogous to the edge states in
the QHE.

\ Following ref.\cite{Mpb} we can now define the spin Hall
conductance
 \bea\label{gsh2}
G^s_{sH}=G_{sH}\frac{\hbar}{2e}=\frac{e}{4 \pi} (N_s-N_t). \eea

This result can be also obtained by calculating the response of
the spin current operator\cite{newref} $$\hat{\bf
J_s}=\frac{\hbar}{4}(\hat{\sigma_z}\hat{\bf v} + \hat{\bf
v}\hat{\sigma_z})$$  to the electric field. This calculation can
easily be done within the framework of the Landauer formalism and
gives the conductance in eq.(\ref{gsh2}). Although this may not be true in
the general case, where $\hat{\bf v}$ does not commute with
$\hat{\sigma_z}$, nonetheless it holds valid in our case, where $\hat{\bf
v}\equiv \frac{\hat{\bf \pi}}{m_e}$ and $[\hat{\bf
\pi},\hat{\sigma_z}]=0$.

\ Now we also want to discuss an interesting effect on the properties
of the reflected current corresponding to the Hall one. In fact
the edge localization of the states implies the presence of rings
of current in the center of the QW see Fig.(1.left). These rings,
with an orbital magnetic momentum $M_z$ are coupled with the spin
($s_z$) so that they  minimize ${\bf M}\cdot {\bf S}$ as shown in
Fig.(4). It follows that also in this case the magnetic properties
of spin up and spin down electrons are opposite.
%%%%%%%%%%%%%%%%%%%%%%%%%%%%%%%%%%%%%%%%%%%%%%%%%%%%%%%%%%%%%%%%%%%%%%%%%%%
\begin{figure}
\includegraphics*[width=1.0\linewidth]{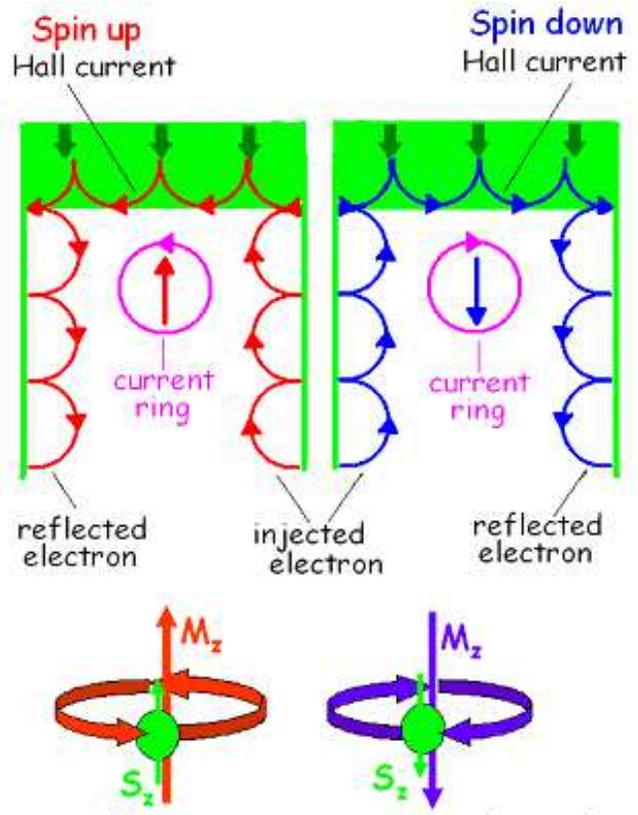}
\caption{(Color online){ Spin and orbital polarization
corresponding to the reflected current. }}
\end{figure}
%%%%%%%%%%%%%%%%%%%%%%%%%%%%%%%%%%%%%%%%%%%%%%%%%%%%%%%%%%%%%%%%%%%%%%%%%%%%
%%%%%%%%%%%%%%%%%%%%%%%%%%%%%%%%%%%%%%%%%%%%%%%%%%%%%%%%%%%%%%%%%

\

{MULTIPROBE AND QPC -} Next, we shortly discuss what happens if we
consider our device as a ballistic four-probe bridge (see
ref.\cite{Mpb} and Fig.(5.left)). In our case we assume that the presence of
transverse currents could be revealed by attaching two leads near
$y=0$. The corresponding currents are $I_1=I_y^0$ and $I_2=I_H$.
Also if we inject a not spin polarized current
($I_1^s=I_1^\uparrow-I_1^\downarrow=0$), when transverse leads are
attached at the boundaries of the QW , pure
($I_2^\uparrow+I_2^\downarrow=0$) spin current
($I_2^\uparrow-I_2^\downarrow\neq 0$) will emerge in the probe 2
of the bridge. The corresponding spin Hall conductance is defined
\cite{Mpb} as in eq.(\ref{gsh}). The correspondence between
the QW and the multiprobe should be discussed in more detail.
The central question is in what follows: if we attach conventional
 Hall probes to a 1D channel,
 in order to measure the Hall voltage, this very procedure destroys the
 1D character of channel at the point where the measuraments are made.
  Thus, a 1D analog
  of  the Hall voltage that can be measured non invasively must be identified.
However it is possible to refer the reader to several papers that
discussed the so called non invasive measurements of the
intrinsic QHE\cite{nim1,nim2} or to other references that proposed
a four terminal measurement of the Hall resistence
in an experimental setup where the QW is connected via Hall probes
to electron reservoirs (Hall contacts) in  the so called  weak link model\cite{wlm1,wlm2}.
\
A different way to observe the QHE is based on the tunneling through a Quantum Point Contact
as discussed by ref.\cite{vw}. This kind of experiment was also proposed for the
study of the non-equilibrium noise in a Chiral Luttinger Liquid i.e. of
the tunneling between edge states in the fractional quantum Hall regime\cite{cham}.
In this case two
 quantum Hall droplets are separated by  a constriction, i.e. the Quantum Point Contact.
Quasiparticles can tunnel across the constriction,
from one edge to the other as we show in Fig.(5.right). As discussed in ref.\cite{cham}
also this experimental setup can be viewed as a four terminal measurement device.
%%%%%%%%%%%%%%%%%%%%%%%%%%%%%%%%%%%%%%%%%%%%%%%%%%%%%%%%%%%%%%%%%%%%%%%%%%%
\begin{figure}
\includegraphics*[width=1.0\linewidth]{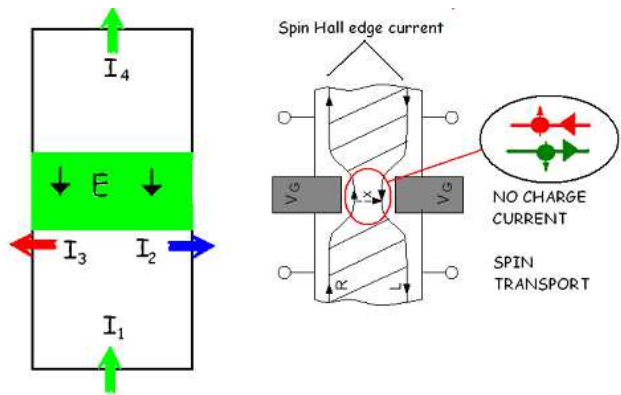}
\caption{(Color online){(Left) The four probe mesoscopic bridge,
for the detection of the pure spin Hall currents, obtained by
attaching tranverse leads to the QW. (Right) Geometries for
tunneling between spin Hall edge states.
 By adjusting the gate voltage  one can obtain either a
simply connected QH droplet  or two disconnected QH droplets.
A pair of electrons  (carrying spin and no charge) can tunnel from one edge to the other.}}
\end{figure}
%%%%%%%%%%%%%%%%%%%%%%%%%%%%%%%%%%%%%%%%%%%%%%%%%%%%%%%%%%%%%%%%%%%%%%%%%%%%
%%%%%%%%%%%%%%%%%%%%%%%%%%%%%%%%%%%%%%%%%%%%%%%%%%%%%%%%%%%%%%%%%

\

{\it DISCUSSION -} Before ending we want to discuss the strength of
the physical quantities in this paper.

Experimentally, in $GaAs-AsGaAl$ interface,  values for $ \alpha e
E_z $ of  order $10^{-11}\; eV\; m$ were observed\cite{Nitta}
corresponding to a triangular potential well of width about $5-10
nm$. It is clear that  the corresponding values for $ \alpha e E_x
$ is smaller  by a factor $L_z/L_x\lesssim 1/5$. Thus if we
consider these kind of devices the effects of the Rashba coupling
are always dominant respect to the ones analyzed in this paper.
Nevertheless,  if the electrons are confined along the $z$
direction in a square well, is possible neglect the Rashba
coupling and our prediction could be tested.

The strength of $E_y$ can be easily by the introduction of
electrodes or Quantum Point Contacts(QPCs)\cite{thor} realized in
split-gate devices. The  width of these devices can be of  the
order of the electron Fermi wavelength and a length much smaller
than the elastic mean free path.

\

{\it CONCLUSIONS -} Here we discuss the theoretical case of a
mesoscopic QW in the ballistic regime by taking in account the SO
coupling effects. We show that the case in the presence of SO
coupling can be reduced to the case where a transverse magnetic
field is present, if we consider spin polarized electrons. We
discuss that in general non commutation implies edge localization
of the currents. This property is the basis for the Spin Hall
Effect and  in this case (ballistic Quantum Wire) it reflects in
the conductance quantization in the longitudinal and transverse
direction. The presence of Hall effect, which we identify as
localization plus reflection, gives also a coupling between rings
of charge current and spin polarization.

%%%%%%%%%%%%%%%%%%%%%%%%%%%%
%%%--------------------------------------------------------

%%%------------------------------------------------------------------------
\bibliographystyle{prsty} %Phys. Rev. style

\bibliography{}

\end{document}